%% file: anonymous-submission-latex-2025.tex
\documentclass[letterpaper]{article} 
\usepackage{aaai25}
\usepackage{times}  
\usepackage{helvet}  
\usepackage{courier}  
\usepackage[hyphens]{url}  
\usepackage{graphicx} 
\usepackage{natbib}  
\usepackage{caption} 
\usepackage{algorithm}
\usepackage{algorithmic}
\usepackage{amsmath}

\usepackage{newfloat}
\usepackage{listings}
\DeclareCaptionStyle{ruled}{labelfont=normalfont,labelsep=colon,strut=off} 
\floatstyle{ruled}
\newfloat{listing}{tb}{lst}{}
\floatname{listing}{Listing}
\usepackage{xcolor}
\usepackage{multirow}

\usepackage{array} 
\usepackage{relsize} 

\title{
Modelling the Spread of Toxicity and Exploring its Mitigation on Online Social Networks}
\author{
    Aatman Vaidya\textsuperscript{\rm 1},
    Harsh Bhagat\textsuperscript{\rm 1},
    Seema Nagar\textsuperscript{\rm 2},
    Amit A. Nanavati\textsuperscript{\rm 1}
}
\affiliations{
    \textsuperscript{\rm 1}School of Engineering and Applied Science, Ahmedabad University\\
    aatman.v@ahduni.edu.in, harsh.b2@ahduni.edu.in, amit.nanavati@ahduni.edu.in

    \textsuperscript{\rm 2} IBM India Research Lab \\
    senagar3@in.ibm.com

}

\begin{document}

\maketitle

\begin{abstract}
Hate speech on online platforms has been credibly linked to multiple instances of real world violence. This calls for an urgent need to understand how toxic content spreads and how it might be mitigated on online social networks, and expectedly has been the topic of extensive research in recent times. Prior work has largely modelled hate through epidemic or spread activation based diffusion models, in which the users are 
often divided into two categories, hateful or not.
In this work, users are treated as transformers of toxicity, based on how they respond to
incoming toxicity. Compared with the incoming toxicity, 
users amplify, attenuate, or replicate (effectively, transform) the toxicity and send it forward. We do a temporal analysis of toxicity on Twitter, Koo and Gab and find that (a) toxicity is not conserved in the network; (b) only a subset of users change behaviour over time; and (c) there is no evidence of homophily among behaviour-changing users. In our model, each user transforms incoming toxicity by applying a {\it shift} to it prior to sending it forward. Based on this, we develop a network model of toxicity spread that incorporates time-varying behaviour of users. We find that the {\it shift}
applied by a user is dependent on the input toxicity and the category. Based on this finding, we propose an intervention strategy for toxicity reduction. This is simulated by deploying peace-bots. Through experiments on both real-world and synthetic networks, we demonstrate that peace-bot interventions can reduce toxicity, though their effectiveness depends on network structure and placement strategy.

\end{abstract}

\input{sections/01_introduction}

\input{sections/02_related_work}
\input{sections/03_analysis}
\input{sections/04_model}

\input{sections/05_intervention}
\input{sections/06_experiments_results}
\input{sections/07_discussion}
\input{sections/08_conclusion}

\bibliography{aaai25}

\clearpage

\input{sections/10_appendix}

\end{document}

%% file: sections/01_introduction.tex
\section{Introduction}
\label{secintro}

Online hate speech has been documented to have produced real-life effects\footnote{https://www.reuters.com/technology/facebook-knew-about-failed-police-abusive-content-globally-documents-2021-10-25/}. The research by \citet{muller2021fanning} demonstrated that anti-refugee sentiment on Facebook directly led to physical attacks against refugees throughout Germany. Hate speech on Facebook led to real world violence in Ethiopia\footnote{https://restofworld.org/2021/why-facebook-keeps-failing-in-ethiopia/}. The Observer Research Foundation identified a direct link between hateful speech online and physical violence in Indian society by tracing cases of mob aggression which started from inflammatory internet content~\cite{mirchandani2018digital}. This underscores the urgent need to understand how hate speech propagates through online social networks (OSN), and to develop effective strategies to mitigate its spread. Of late, the study of hate speech and its spread is being examined through multiple perspectives and approaches in order to gain insights into this problem. The expectation from these efforts is to (a) understand the process of spreading of hate and (b) find ways of mitigating it. \\

Prior research has approached this problem from multiple angles. One line of work classifies messages or users, as hateful vs. non-hateful, in order to study user behaviour changes and the implications of the structure of the network connecting them~\cite{ribeiro2018characterizing, Mathew2018SpreadOHA}. Another (often overlapping) line of work attempts to analyse and model the flow of hate through the network. Several variants of Belief propagation~\cite{Mathew2018SpreadOHA} and Spreading and Activation (SPA) based models~\cite{nagar2021capturing} have been used to model the diffusion of hatred through social networks. Other works have extended epidemic spread models for this purpose as well~\cite{yousefi2024comparative}.

We presuppose that hate is not binary. We use {\it toxicity} to quantify the degree of hatred in a message~\cite{perspectiveapi}. Given an underlying network, each user receives an incoming toxicity in the range [0-1] and outputs an outgoing toxicity in the range [0-1]. Each user is thus a transformer of toxicity. A user responds to a stimulus of the input toxicity they receive by applying a {\it shift} to it. This leads us to investigate the following questions:
\begin{description}
    \item \textbf{RQ1:} How does toxicity spread in the network? Does it depend on the structure of the network?
    \item \textbf{RQ2:} Are some users more responsible for the spread of toxicity than others? Does user behaviour change with time? 
    \item \textbf{RQ3:} Are there ways to mitigate the spread of toxicity? Especially, are there soft interventions which do not require the removal of users or connections to achieve this?
\end{description}

\begin{figure}[h]   
\centering
\includegraphics[width=\columnwidth]{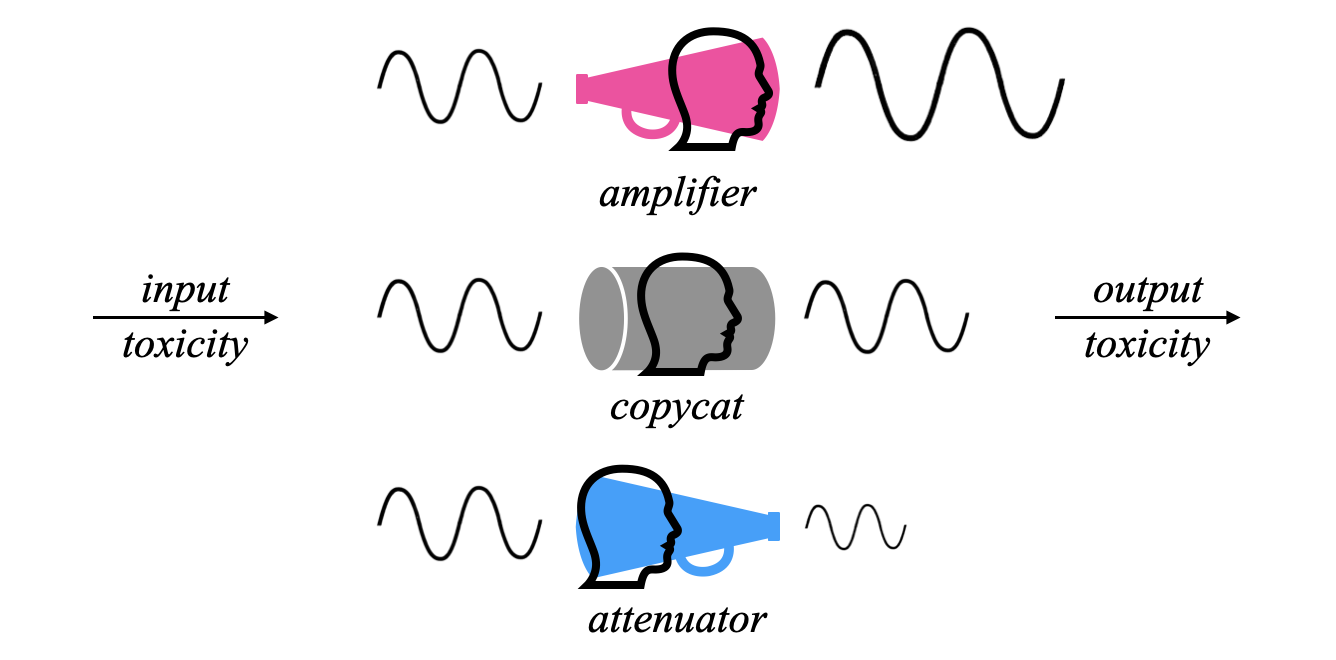}
\caption{Users viewed as transformers of toxicity: amplifiers, users whose output toxicity is higher than their input toxicity; copycats, whose output toxicity is almost the same as their input toxicity; and attenuators, whose out toxicity is less than their input toxicity.}
\label{fig:xformers}
\end{figure}
\begin{figure}[h]
\centering
\includegraphics[width=\columnwidth]{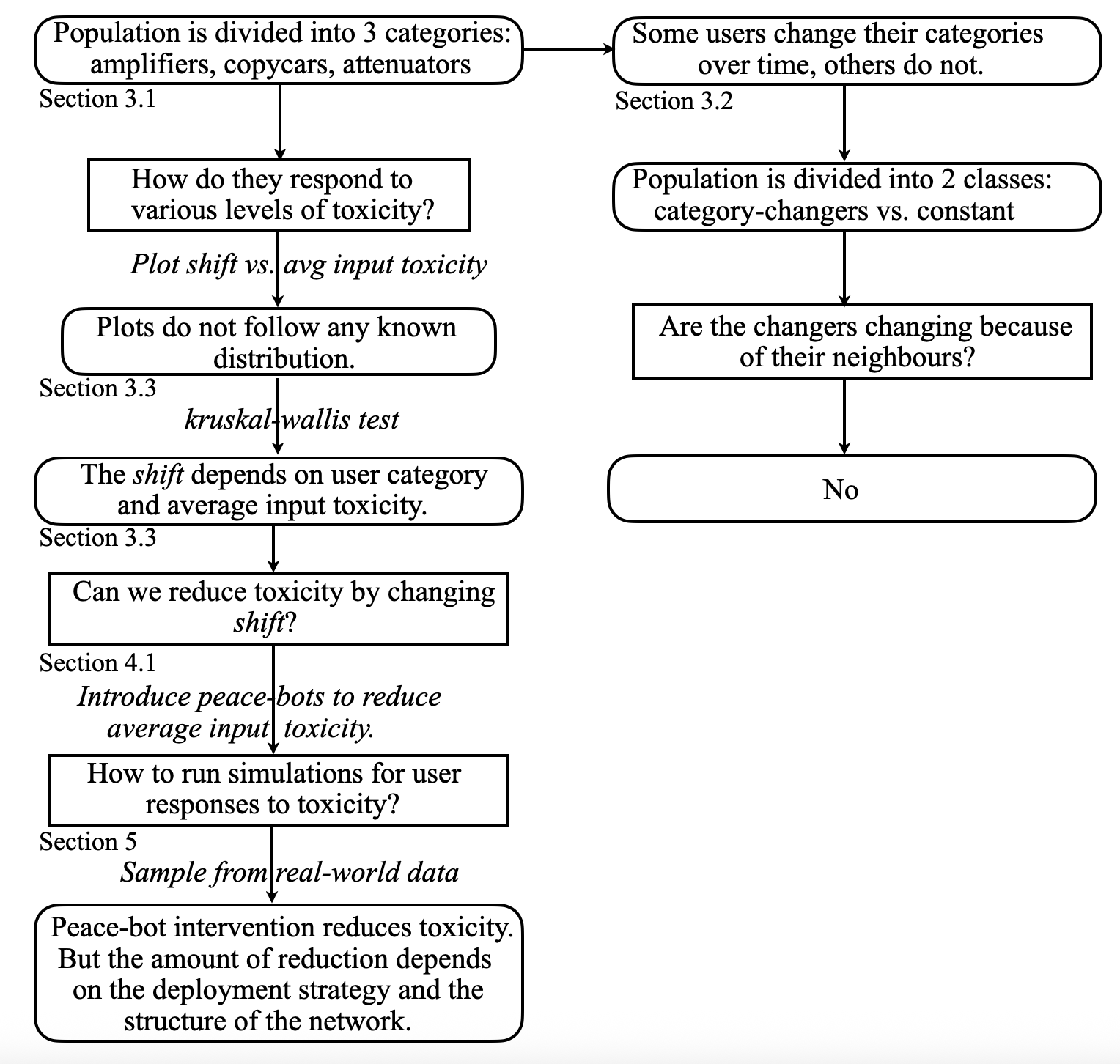}
\caption{The flow of questions explored in the paper, beginning with the understanding that there are three categories of users.}
\label{fig:paper-flow}
\end{figure}

To answer these questions, we experimented on three large scale online social networks: Twitter, Koo and Gab, and analysed them for the spread of toxicity over time. Figure~\ref{fig:paper-flow} describes the flow of this paper. Our contributions in this work are the following:
\begin{itemize}
    \item Based on empirical findings, we classify users into three distinct categories based on how they respond to toxicity. \textit{Amplifiers} (who increase it), \textit{Attenuators} (who decrease it), and \textit{Copycats} (who propagate it with little change), see Figure~\ref{fig:xformers}. We show that this categorization is observable across all three platforms.
    \item Through a temporal analysis, we show that standard epidemiological models don't adequately capture toxicity spread. We find that nearly half of the users remain in a fixed user category over time, while others fluctuate. This is a phenomenon not captured by traditional Susceptible-Infected-Recovered (SIR*) like frameworks.
    \item We develop a new model for toxicity spread based on our findings that a user applies a ``\textit{shift}'' in toxicity depending on their category and the level of toxicity they receive.
    \item Based on this model, we propose a soft intervention using peace-bots to strategically lower the average toxicity users are exposed to, hence, decreasing the total toxicity observed in the social network.
    \item We conduct simulations on both real-world and synthetic networks to evaluate our model and peace-bot strategy. Our experiments show that most effective deployment strategy of peace-bots is highly dependent on the underlying network structure, with different strategies proving optimal for different platforms.
\end{itemize}

%% file: sections/02_related_work.tex
\section{Related Work}
\label{sec:related}

\paragraph{Modelling Spread of Hateful Content.}Several studies have examined how hateful or toxic content spreads through online social networks.~\citet{Mathew2018SpreadOHA} used belief propagation to model the diffusion of hateful content, showing that it travels farther, faster, and reaches a wider audience than non-hateful content. In a follow-up study,~\citet{Mathew2019HateBHA} analysed hate speech on Gab using a DeGroot model, finding that hateful users tend to become central quickly, and that newer users adopt hateful behaviour faster.~\citet{ribeiro2017like} studied the user characteristics of hateful users on Twitter, and highlighted that hateful accounts differ from normal users in activity, network centrality, and content type.

Recent work has explored more nuanced modelling approaches.~\citet{vaidya2024analysing} proposed a model of toxicity spread on Twitter, classifying users into behavioural categories based on their responses to incoming content.~\citet{lerman2024affective} examined affective polarization, showing that toxicity-driven polarization is not limited to group-based divides but is instead a structural property of social networks.~\citet{Maarouf2022TheVOA} analysed virality, finding that hateful content from verified users is disproportionately more likely to spread widely.

Topic based modelling approaches have also been proposed. ~\citet{nagar2021hate} developed a graph autoencoder that integrates user and textual features to track the spread of hate.~\citet{goel2023hatemongers}, consistent with~\citet{Mathew2018SpreadOHA}, found that hateful users play a more crucial role in governing the spread of information compared to singled-out hateful content. They also observe that hatemongers dominate the echo chambers in a network.~\citet{masud2021hate} introduced a topic-aware diffusion model with attention mechanisms that leverages news data to predict hateful retweets, while ~\citet{gupta2021consumption} combined topic modelling and ensemble classification to show that hateful tweets receive more retweets than non-hateful ones.

Another prominent line of work applies epidemiological models to capture the spread of hate and toxicity.~\citet{obadimu2020developing} introduced the STRS model (Susceptible–Toxic–Recovered–Susceptible) on YouTube, showing that toxicity can escalate into an epidemic if the reproduction number $R_0 > 1$, and highlighting the importance of interventions that reduce exposure.~\citet{addai2024seiqr} adapted the SEIQR model on Twitter, incorporating user history and index of memory can provide a nuanced modelling. ~\citet{maleki2022applying} applied the SEIZ model (Susceptible–Exposed–Infected–Skeptic) to study toxicity propagation on Twitter, demonstrating its effectiveness compared to alternative epidemiological models.~\citet{dagtas2024modeling} evaluated five epidemiological models on Reddit data, achieving high accuracy with less than 2\% fitting error. Building on this,~\citet{yousefi2024developing} extended epidemic models to differentiate between moderate and highly toxic users, yielding improved predictive performance.

\paragraph{Homophily} also described as users being ``birds of a feather flock together'' highlights that users with similar traits are more likely to connect~\cite{mcpherson2001birds, kossinets2009origins}.~\citet{halberstam2016homophily} measured homophily among politically engaged users on Twitter, and found that majority group users have more connections, greater information exposure, and faster access to information than minority group users, showing the tendency for like-minded content to circulate within groups.

\paragraph{Bots on Online Social Networks.}Social bots have played a key role in spreading digital content on social networks~\cite{ferrara2016rise, morgan2018fake}. Social media bots are defined as automated accounts that engage in content creation, distribution, and collection~\cite{ng2025social}. Empirical studies show that bots can shape information dynamics in important ways. For instance,~\citet{monsted2017evidence} demonstrated through controlled bot experiments that diffusion on Twitter follows complex contagion dynamics rather than simple contagion.~\citet{shao2018spread} found that bots amplify articles from low-credibility sources during the early stages of diffusion, helping such content gain traction.~\citet{Uyheng2020BotsAO} linked bot activity to increased levels of hate, particularly in dense and isolated communities in a network.

%% file: sections/03_analysis.tex
\section{Temporal Analysis}
\label{sec:temporal}

\subsection{Datasets and Pre-processing}


In this section, we describe the three datasets used in our analysis to understand the temporal dynamics of toxicity spread on social networks. Each dataset contains a directed graph $G$, the structure and edge type of the graph vary across datasets. All datasets include a collection of text-only posts gathered over time from multiple users\footnote{The Koo and Gab datasets are publicly available under the CC-By Attribution 4.0 licence.}. (See Table~\ref{tab:dataset_comparison_vertical}).

\noindent \textbf{(1) Twitter}: The dataset is published by \cite{ribeiro2017like} and presents a Twitter dataset of $100K$ users along with up to 200 tweets from their timelines with a random walk-based crawler on the retweet graph ($G$), and select a subsample of $4, 972$ to be manually annotated as hateful or not through crowdsourcing with information about hateful tweets and users. The authors also look at  user activity patterns of hateful and normal users.

\noindent \textbf{(2) Gab}: The dataset is created by \cite{doi:10.1073/pnas.2212270120} and studies prevalence of fear and hate speech on the social network. It contains $700K$ hateful and $400K$ fear speech posts collected from Gab.com.  The dataset contained information about posts reshared by users, and we constructed a directed graph based on these reposts. 

\noindent \textbf{(3) Koo}:~The dataset is published by \cite{mekacher2024koo} and presents largest publicly available Koo dataset, spanning from the platform’s founding in early 2020 to September 2023, containing $72M$ posts, $40M$ shares and other metadata. Koo was a popular microblogging platform based in India. Similar to Gab, the dataset contained information about posts re-shared by users and we constructed a directed graph based on shares.\\

For each dataset, we assign a toxicity score using Perspective API to every post~\cite{perspectiveapi}. Toxicity is defined as a rude, disrespectful, or unreasonable comment that is likely to make people leave a discussion. 

We wanted to examine how toxicity evolves over time in a network and how users respond to toxicity. To do so, we filtered all three datasets to find a set of users who were present during a time period. Depending on the dataset, the temporal unit of the time period was either weeks or months. Table~\ref{tab:dataset_comparison_vertical} provides details of the selected timelines and the corresponding user statistics.

\begin{table}[h]
\centering
{\relsize{-0.32}
\setlength{\tabcolsep}{2.7pt}
\setlength{\extrarowheight}{1pt}
\begin{tabular}{cccrrr}
\hline
\textbf{Dataset} & \textbf{Timeline}                                           & \textbf{\begin{tabular}[c]{@{}c@{}}Time\\ Unit\end{tabular}} & \textbf{\begin{tabular}[c]{@{}c@{}}No. of \\ Posts\end{tabular}} & \textbf{Nodes} & \textbf{Edges} \\ \hline
\multicolumn{6}{l}{\hspace{0.5in}\textit{Complete Dataset}} \\ \hline
Twitter          & 01/2017--10/2017 & Week                                                         & 17.2M                                                            & 99.8K          & 2.27M          \\ \hline

Gab              & 10/2016--06/2018 & Month                                                        & 20.1M                                                            & 62.3K          & 10.4M          \\ \hline
Koo              & 01/2020--09/2023& Month                                                        & 16.9M                                                            & 214.9K         & 1.93M          \\ \hline
\multicolumn{6}{l}{\hspace{0.5in}\textit{Temporal Subset}}                                                                                                                                                                                                               \\ \hline
Twitter          &07/2017--10/2017& Week                                                         & 1.34M                                                            & 40.9K          & 737K           \\ \hline
Gab              &11/2017--06/2018& Month                                                        & 6.08M                                                            & 53.2K          & 2.19M          \\ \hline
Koo              &08/2022--03/2023& Month                                                        & 2.96M                                                            & 49.1K          & 1.14M          \\ \hline
\end{tabular}
}
\caption{Datasets Overview and Temporal Subsets.}
\label{tab:dataset_comparison_vertical}
\end{table}

In the next section, we list our findings from analysing toxicity spread across the three datasets. 

\subsection{Preliminary Analysis}

\begin{figure}[h]
\centering
\includegraphics[width=\columnwidth]{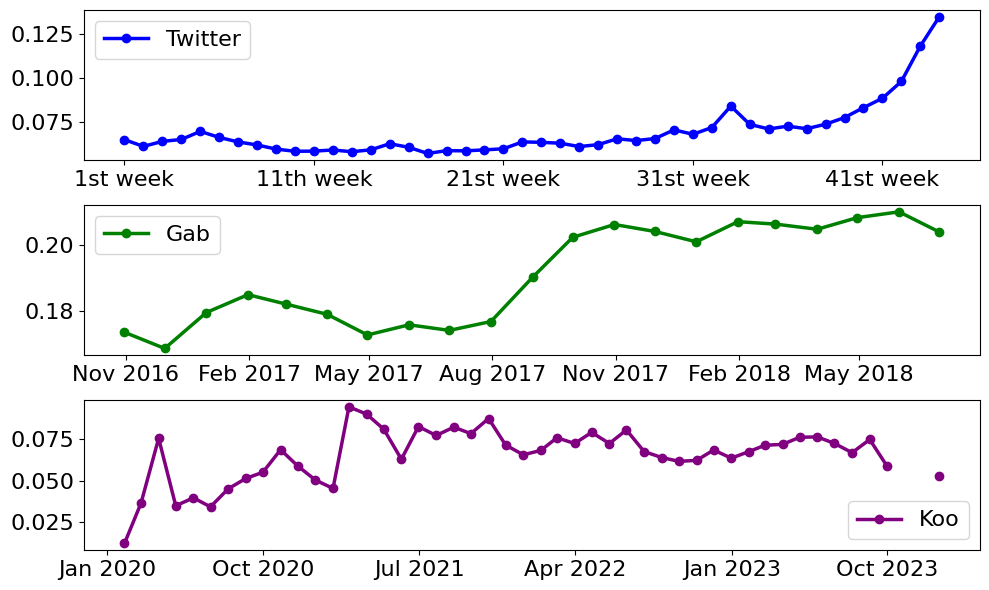}
\caption{Average Toxicity over time in all the datasets.}
\label{fig:total_tox_koo}
\end{figure}

We begin by examining how toxicity changes over time. First, we analyse the Twitter dataset and then repeat similar experiments on the Gab and Koo datasets. Figure~\ref{fig:total_tox_koo} shows the average toxicity changes across time on all three datasets. \citet{nagar2021capturing} used spread activation (SPA) models to capture toxicity in networks, however, our analysis in Figure~\ref{fig:total_tox_koo} indicates that these models don't adequately capture its spread.

To better understand how users react to toxicity from a network perspective, we investigated whether a user’s neighbourhood plays a role in influencing their behaviour. 
To measure the influence of a user $u$’s neighbourhood on $u$, we calculate the difference between $u$’s average toxicity and the average toxicity of $u$'s in-degree neighbours over time.
We call this difference \textbf{shift}. A shift represents the average change a user applies to the incoming toxicity before transmitting the message further. 

\begin{figure}[h]
\centering
\includegraphics[width=\columnwidth]{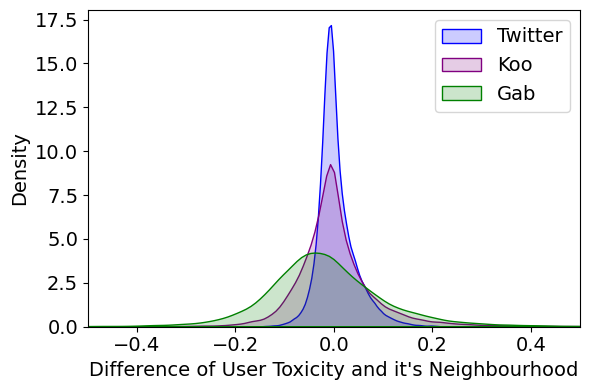}
\caption{Distribution of the difference between User's Average Toxicity and the Average Toxicity of its in-degree neighbourhood in all three datasets. All the distribution's fail the Shapiro–Wilk (SW) test for normality.}
\label{fig:dist_diff_tox_koo}
\end{figure}

Figure~\ref{fig:dist_diff_tox_koo} shows the distribution of shifts in all the three datasets. We then applied the Shapiro–Wilk test to check for normality of these distribution's, and the results showed that is not a normal distribution~\cite{shapiro1965analysis}.

\begin{figure}[h]
\centering
\includegraphics[width=\columnwidth]{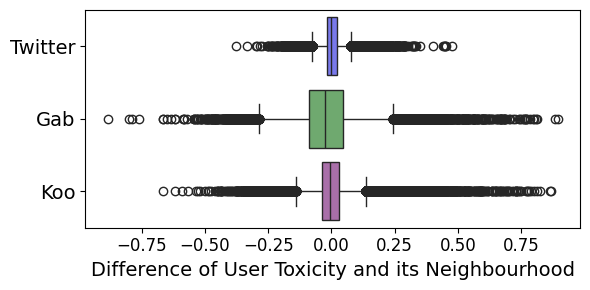}
\caption{Box and Whisker plot for the distributions in Figure~\ref{fig:dist_diff_tox_koo}. Since the data is not normal, we use the IQR
method to detect outliers and separate them from normal
users. The outlier users on the right are called \textit{amplifiers}, on
the left are called \textit{attenuators} and the remaining are called \textit{copycats}.}
\label{fig:iqr_dist_diff_koo}
\end{figure}

In such a case where the distribution is not normal, we used the Interquartile Range (IQR) measure to categorise users based on their \textit{shift} behaviour. This helps detects outliers in the distribution and separate users. Figure \ref{fig:iqr_dist_diff_koo} shows the box-and-whisker plot of the distribution in \ref{fig:dist_diff_tox_koo}. This approach naturally divides the set of users into 3 disjoint subsets and help us categorise user behaviour~\cite{vaidya2024analysing}. These categories reflect how users respond to toxicity. The outlier users on the right in Figure~\ref{fig:iqr_dist_diff_koo} are called \textit{amplifiers}, on the left are called \textit{attenuators} and the remaining (typical users) are called \textit{copycats}. The amplifiers send out more toxicity than they receive, the attenuators send out less toxicity than they receive and the copycats send out almost the same toxicity as they receive (see Figure~\ref{fig:xformers}).\\

\citet{vaidya2024analysing} employed the same methodology and showed that average toxicity in the network changes over time. They further demonstrated that the placement of amplifiers, attenuators, and copycats influences the extent of toxicity spread. While the authors analysed the dataset in aggregate, this does not preclude the possibility that user behaviour changes over time. If toxicity is viewed as a disease, one may ask whether users exhibit temporary symptoms of increasing/decreasing it (i.e., becoming amplifiers/attenuators), and by default, all users are copycats. Prior work has explored epidemiological models such as SIR and SIER to study the diffusion of hate~\cite{yousefi2024comparative, maleki2022applying, dagtas2024modeling}. A temporal analysis can therefore provide valuable insights into whether such approaches offer an effective framework for modelling hate.

\begin{figure}[h]
\centering
\includegraphics[width=\columnwidth]{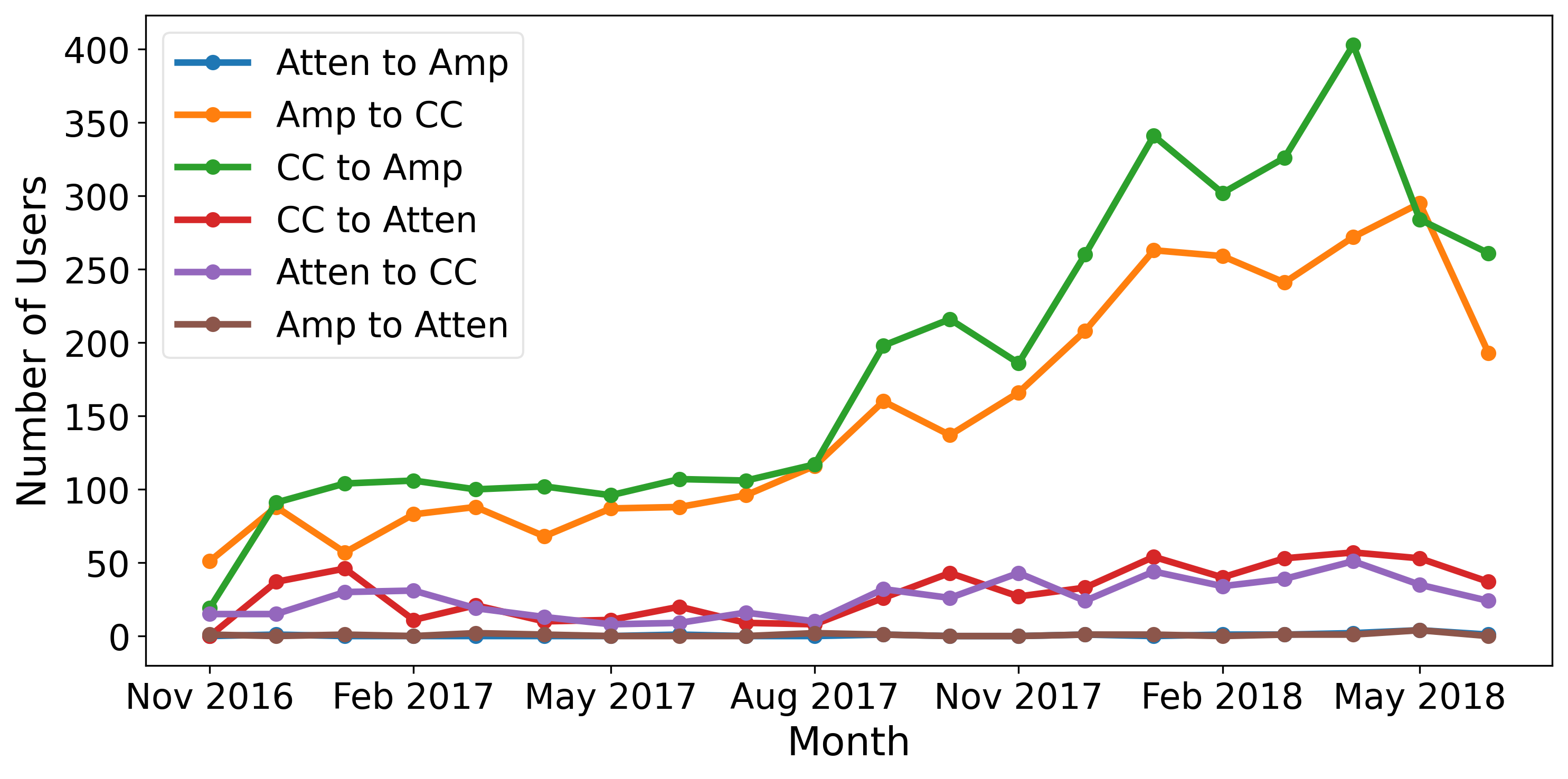}
\caption{Change in User Category in the Gab Dataset. We observed that a small set of users' change behaviour. Similar Figures for Twitter and Koo have been added in Appendix~\ref{sec:appendix_preliminary_analysis}.}
\label{fig:user_category_change_over_time_gab}
\end{figure}

Hence, we looked at how users change behaviour over time. Figure~\ref{fig:user_category_change_over_time_gab} shows that (i) not all users are consistently present in every time interval, (ii) only a set of users change behaviour across time, (iii) users do not necessarily experience all categories, and their transitions show no fixed patterns. This is unlike diseases, where individuals typically transition between well-defined states (susceptible, infected, recovered). This suggests that epidemiological models such as SIR may have limited effectiveness in fully capturing the spread of toxicity on a network.

An immediate question that follows is, {\it why do some users change their behaviour?} To check if users change their behaviour because of their neighbours, we calculated the network-level homophily of users~\cite{easley2010networks}. We defined users who changed their behaviour over time as \textit{changing} users and those who did not as \textit{non-changing} users. The results, presented in Table~\ref{tab:homophily-changing-nonchanging}, show that there is no evidence of homophily within the set of changing users, within the set of non-changing users, or between the two groups.

\begin{table}[h]
\centering
\begin{tabular}{ccc}
\hline
\textbf{User Category} & \textbf{Actual Edges $(x)$} & \textbf{Probability $(x^2)$} \\ \hline
changing (p)           & 0.012234              & 0.008665                   \\ \hline
not changing (q)       & 0.869270              & 0.822492                   \\ \hline
\end{tabular}
\caption{Homophily between Changing and Non Changing users in the Network. Actual edges $(x)$ represent the fraction of edges within a given category to the total number of edges in the graph. The probability $(x^2)$ is the squared value of actual edges. A user category is said to exhibit network homophily if $x$ is much larger than $x^2$.}
\label{tab:homophily-changing-nonchanging}
\end{table}

We next ask,~{\it how do users in the three categories respond to incoming toxicity?} How does their output toxicity change with incoming toxicity? In order to understand this, we inspect the shift distributions in the next section, of all categories of users across all three datasets. 

\subsection{Shift Analysis}

\begin{figure*}
    \centering
    \includegraphics[width=0.8\linewidth]{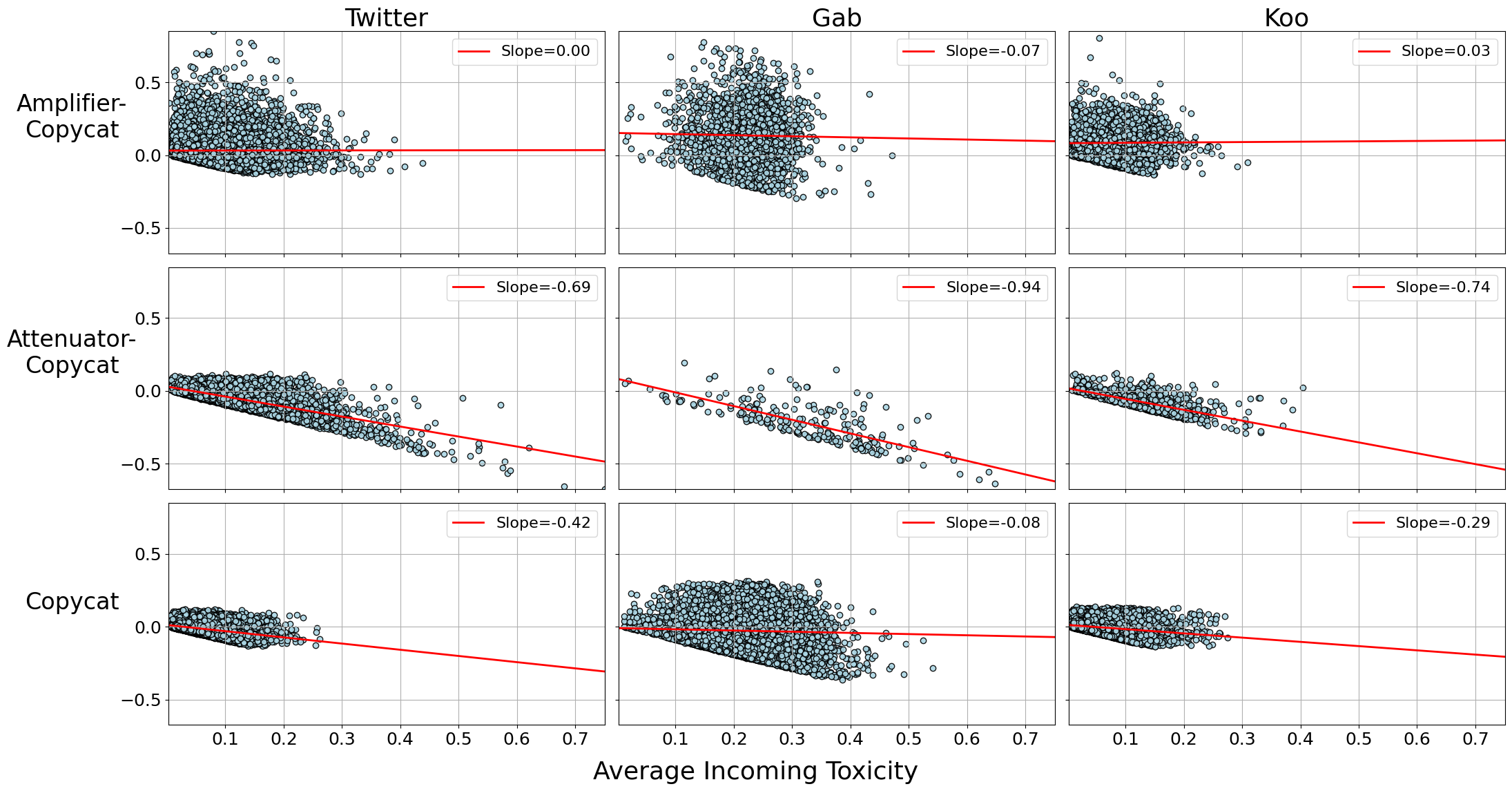}
    \caption{Distribution of shifts as a function of Average Incoming Toxicity across all datasets. Average incoming toxicity for a user $u$ is the average toxicity value of all in-degree neighbours of $u$. Given that most users exist in only one of copycat$\leftrightarrow$amplifier, copycat$\leftrightarrow$attenuator, and copycat categories, we plotted the distribution so that each user's behaviour is reflected in one of the plots only.}
    \label{fig:summary-pic-1}
\end{figure*}

A key aspect to understanding user behaviour is, how do different categories of users respond to different ranges (say low, medium, high) in input toxicity? Is their reaction consistent regardless of the input range? If not, then how does the variation differ across the three categories of users? These questions lead us to inspecting the shift behaviour.

For this analysis, we require a timeline with a consistent set of users. Hence, we filter each dataset to retain such subsets of users. For Twitter we analyse 7,170 users and their neighbourhoods over 13 weeks; for Gab, 4,808 users over 8 months; and for Koo, 5,645 users over 8 months. Further details about these temporal subsets are provided in Table~\ref{tab:dataset_comparison_vertical}, where the reported totals include both users and their neighbourhoods.

\begin{table}[h]
\centering
\begin{tabular}{cccc}
\hline
\textbf{Dataset} & \textbf{Amplifier} & \textbf{Attenuator} & \textbf{Copycat} \\ \hline
Twitter          & [0, 0.50]          & \textbf{[0, 0.80]}    & [0,0.3]          \\ \hline
Gab              & [0, 0.50]          & \textbf{[0, 0.70]}    & [0,0.6]          \\ \hline
Koo              & [0, 0.35]          & [0, 0.40]           & [0,0.3]          \\ \hline
\end{tabular}
\caption{Range of Average Incoming Toxicities experienced by each category.}
\label{tab:shift-1}
\end{table}

\begin{table}[h]
\centering
\begin{tabular}{cccc}
\hline
\textbf{Dataset} & \textbf{Amplifier} & \textbf{Attenuator} & \textbf{Copycat} \\ \hline
Twitter          & [ 0.0, 0.9]        & [-0.7, 0.20]        & [-0.20,0.20]     \\ \hline
Gab              & [-0.2, 0.8]        & [-0.7, 0.20]        & [-0.40,0.30]     \\ \hline
Koo              & [-0.1, 0.8]        & [-0.3, 0.15]        & [-0.15,0.15]     \\ \hline
\end{tabular}
\caption{Range of the shifts applied by each category of users.}
\label{tab:shift-2}
\end{table}

Figure~\ref{fig:summary-pic-1} shows the shift distributions of users in each category for all datasets in temporal subsets.   
We summarise some of the information in Tables~\ref{tab:shift-1} and \ref{tab:shift-2}. We make the following observations:
\begin{itemize}
\item (Table~\ref{tab:shift-1}) The range of average input toxicities experienced by each category of users is different. The attenuators in Twitter and Gab seem to experience higher inputs toxicities compared to the rest of the users.
\item (Table~\ref{tab:shift-2}) The range of the shifts applied by each category of users is different.
Expectedly, the amplifiers' positive shifts are the largest and the attenuators' negative shifts are the largest.
\end{itemize}

For each type of user category, we checked if the distribution of shifts by input toxicity follows any known distribution. We tested for normality, power-law, lognormal, exponential, stretched-exponential and truncated-power-law, and none of them passed~\footnote{https://github.com/jeffalstott/powerlaw}~\cite{alstott2014powerlaw}. 
Given that this data does not follow any of the popular named distributions, we conducted the $k$-sample Anderson-Darling test~\cite{scholz1987k} to check if the various shifts at least belong to the same (un-named) distribution. We found that no two of them belong to the same distribution either. 

\begin{table}[]
\centering
\begin{tabular}{cccc}
\hline
\textbf{Dataset} & \textbf{\begin{tabular}[c]{@{}c@{}}Category\\ Effect\end{tabular}} & \textbf{\begin{tabular}[c]{@{}c@{}}Predecessor \\ Effect\end{tabular}} & \textbf{\begin{tabular}[c]{@{}c@{}}Interaction \\ Effect\end{tabular}} \\ \hline
Twitter          & 0.0000                                                             & 0.0363                                                                 & 0.0631                                                                 \\ \hline
Koo              & 0.0000                                                             & 0.0000                                                                 & 0.0000                                                                 \\ \hline
Gab              & 0.0000                                                             & 0.4745                                                                 & 0.4595                                                                 \\ \hline
\end{tabular}
\caption{$p$-values of Kruskal–Wallis test.}
\label{tab:kw-test}
\end{table}

Then we conducted the Kruskal-Wallis test~\cite{kruskal1952use} to check if the shift applied by a user to average incoming toxicity depends upon (a) the user category and (b) the value of incoming toxicity. We found that for the Twitter and Koo datasets, it depends on both; while for Gab, it depends only on the user category. We surmise that the reason for this is that the Gab dataset was created to specifically study hate speech and thus contains mostly hateful posts (see Table~\ref{tab:kw-test}). The dependence of the shift on these two factors forms the basis of a model for the spread of toxicity, which we discuss in the next section.

%% file: sections/04_model.tex
\section{The Model}
\label{sec:model}

From the analysis, we found that a user's response to average incoming toxicity is not static. The shift applied by a user to average incoming toxicity depends upon (a) the user category and (b) the value of incoming toxicity. At each time step $t$, the output toxicity of a user $u$ is governed by:
\begin{eqnarray}
O(u,t) = I_{\textit avg}(u,t) + s(c(u,t),I_{\textit avg}(u))
\label{eqn:model}
\end{eqnarray}
where $O(u,t)$ is the output toxicity, $I_{\textit avg}$ is the average of the input toxicity, $c(u)$ is the category of the user and $s$ is the shift applied by the user. The general model above can be suitably tailored for datasets such as Gab where the dependence on incoming toxicity is not observed.\\

From the analysis, we found that the distributions of the shifts as a function of input toxicity do not fit any well known distributions. Due to this, we sample shifts from the Twitter distribution (see Figure~\ref{fig:summary-pic-1}) for all the experiments in this paper.

\begin{algorithm}[h]
\caption{{\it tox\_spread}}
\label{alg:model}
\begin{algorithmic}[1]
\REQUIRE The network of users $G=(V,\vec{E})$, shift distributions $D_{\textit amp}, D_{\textit atn}, D_{\textit cc}$ for each user category.
\REQUIRE The number of iterations, {\it kiter}.
\REQUIRE The initial tweet(s) with their toxicities. 
\ENSURE The final values of toxicity for each user.  
\FOR{each timestamp $t$ in $[1,\ldots,{\it kiter}]$}
\FOR{node $u \in V$, $\textit{tox}(u)\not = 0$}
\FOR{$v \in out(u)$ // $v$ has an incoming edge from $u$}
\STATE {\it currcat = catg(v)} //  {\it catg}$v$ is one of {\it amp, atn, cc}
\STATE {\it avgintox(v) = $\frac{1}{|in(v)|}\sum_{v\in out(u))} tox(u)$}
\STATE sample {\it shift} from $D_{\textit currcat}$ given {\it avgintox(v)}
\STATE {\it tox(v) = avgintox(v) + shift} 
\IF{$\textit{tox(v)} > 1$}
\STATE $\textit{tox(v)} \gets 1$
\ELSIF{$\textit{tox(v)} < 0$}
\STATE $\textit{tox(v)} \gets 0$
\ENDIF
\STATE update the category of $v$ // may remain unchanged
\ENDFOR
\ENDFOR
\ENDFOR
\end{algorithmic}
\end{algorithm}

We now formulate a model for the spread of toxicity, detailed in Algorithm~\ref{alg:model}. We specify the following assumptions and model parameters:
\begin{itemize}
    \item The model process is initialized with one or more starting nodes, each containing a message with initial toxicity $> 0$.
    \item Time is represented as a sequence of forward passes (hops), where a hop denotes the action of a user forwarding a message to its successors (out-degree neighbours).
    \item A fraction of users change category with time.
    \item Shifts are sampled from real-world dataset distributions.
\end{itemize}

We detail the model in Algorithm~\ref{alg:model}:{\it tox\_spread}. For each user $u$ with one or more incoming edges from nodes carrying non-zero toxicities, we compute the average input toxicity. Based on this value and the category of $u$ at that time step, we sample from the real-world distribution to determine the shift applied by $u$. 


%% file: sections/05_intervention.tex
\subsection{Intervention}
\label{sec:intervene}

Given a model that captures the spread of toxicity, a natural next question is how it might be mitigated? Prior work has explored approaches such as removing hateful nodes or links from the network to limit the spread of hate and toxicity~\cite{Alorainy2022DisruptingNOA, Artime2020EffectivenessODA}. Instead, we focus on a gentler approach. Since the shift applied by a user depends on the average incoming toxicity, can we alter the {\it average} incoming toxicity such that the output toxicity reduces?

Intuitively, since most users are copycats, {\it reducing} the average incoming toxicity ought to lead to an overall reduction in toxicity.
Suppose we deploy ``peace-bots", i.e., bots which send out non-toxic messages (with toxicity = 0). This would lead to a reduction in the average incoming toxicity of the outgoing neighbours of the peace-bots. Several questions arise: 
\begin{itemize}
    \item \textbf {Q1}: How can we be assured that the toxicity will reduce? How does this depend on the shape of the shift distributions $D_{\textit amp}, D_{\textit atn}, D_{\textit cc}$?
    \item \textbf {Q2}: How many peace-bots do we need? 
    \item \textbf {Q3}: How to decide where to deploy them in the network? 
\end{itemize}

In order to answer Q1, let us consider the distributions and Equation~\ref{eqn:model}. 
Rewriting the equation (the parameters have been dropped from the equations for brevity):
\begin{eqnarray*}
O &=& I_{\textit avg} + s(c,I_{\textit avg}) \\
O' &=& I_{\textit avg}' + s'(c,I_{\textit avg}') \\
\end{eqnarray*}
The latter equation with $I_{\textit avg}',s',O'$ denotes the values after the intervention (i.e. after the deployment of peace-bots). 
So, $I_{\textit avg}' < I_{\textit avg}$.
For the intervention to be beneficial, $O' < O$, implying that, $I_{\textit avg}' + s' < I_{\textit avg} + s$.


Therefore, in the shift distribution $D$, 
{\it if the expected value of the output toxicity is smaller for smaller average input toxicities, then the peace-bot intervention can be expected to work}.







To answer Q2, Q3 and to check how effective peace-bot intervention actually is, we turn to experiments.

%% file: sections/06_experiments_results.tex
\section{Experiments and Results}
\label{sec:results}
We evaluate the proposed model and intervention strategy on both randomly generated and real-world networks. We conduct extensive experiments, varying both the number of peace-bots and their positions within the network.

\subsection{Experimental Setup}

We considered random graph models like Barabási–Albert (BA), Erdős–Rényi (ER) graph and Watts–Strogatz (WS)~\cite{barabasi1999emergence, erdHos1961strength, watts1998collective}. For our experiments, we select ER graphs because BA and WS graphs are less representative of real-world social media networks~\cite{chang2025llms}. For instance, WS graphs fail to reproduce realistic degree distributions, while BA graphs capture power-law degree distributions but lack community structure and clustering. To address these limitations, we also evaluate our model on real-world graphs of Twitter, Koo, and Gab, as summarized in Table~\ref{tab:dataset_comparison_vertical}. We aim to address the following questions through our experiments:
\begin{itemize}
    \item \textbf{Q1}: Does Intervention strategy of peace-bots help reduce toxicity in the network?
    \item \textbf{Q2}: How does the number of peace-bots matter?
    \item \textbf{Q3}: Does the placement of peace-bots matter?
    \item \textbf{Q4}: Does network structure matter?
\end{itemize}

We use the following parameters for our experiments:
\begin{itemize}
    \item \textit{Graph Sizes:} We generate directed ER Graphs $(G_{n,p})$ of sizes $25K, 50K, 75K$ and $100K$ nodes with a probability $p$ of $0.05\%$ for edge creation. We also test the model on real-world graphs with sizes shown in Table~\ref{tab:dataset_comparison_vertical}.
    \item \textit {Timeline:} To simulate time in the model, we consider 3--4 hops as a week and run the model for 8 weeks. (24--32 hops) (see definition of time and hops in the Algorithm~\ref{alg:model})
    \item \textit {User Distribution:} We keep the proportion of Amplifiers 5.3\%, Attenuators 1.4\% and Copycats 93.3\%. We randomly assign all the nodes in the graph to these user proportions. Along with this, 47\% users change their category with some probability detailed in Table \ref{tab:user_category_change_prob} in Appendix. The numerical values chosen are observed in the Twitter dataset. 
    \item \textit {Shift Distribution:} To sample shift values, we draw from the distributions observed in the real-world datasets (see Figure~\ref{fig:summary-pic-1}). For a given value of input toxicity, we consider the corresponding range of shift values and apply a density-based probability sampling approach to select the shift. For our experiments, we chose the shift distribution from the Twitter dataset.
\end{itemize}

For every week in the simulation, we record total toxicity and average toxicity per user in the network. For each graph size, we run the experiment five times and take average of the total toxicity to find the results. We ran all the experiments on a Mac Studio with 96GB of unified memory. 

\subsection{Results}

To understand the placement of peace bots and its effect on total toxicity we devise the following scenarios to run the experiments:
\begin{enumerate}
    \item Case 1 -- No peace-bots assigned.
    \item Case 2 -- $N$ peace-bots assigned as indegree neighbours to randomly selected nodes. We call this the Random Placement (RP) strategy.
    \item Case 3 -- $N$ peace-bots assigned as indegree neighbours to nodes with the lowest indegree. We call this the Lowest Indegree (LI) strategy. 
\end{enumerate}

To measure the effect of peace-bots in reducing the total toxicity, we tabulate the {\it percentage reduction} in the toxicities based on the RP and LI strategies, with the Case 1 as the baseline.\\

\begin{table}[h]
\centering
\setlength{\extrarowheight}{1pt} 
\begin{tabular}{|r|r|r|r|r|}
\hline
\textbf{Nodes}        & \textbf{Edges}         & \textbf{\begin{tabular}[c]{@{}c@{}}No. of \\ Peace Bots\end{tabular}} & \multicolumn{1}{c|}{\textbf{RP}} & \multicolumn{1}{c|}{\textbf{LI}} \\ \hline
\multirow{6}{*}{25K}  & \multirow{6}{*}{312K}  & 280                                                                   & 2.84\%     & 3.26\%     \\ \cline{3-5} 
                      &                        & 560                                                                   & 3.58\%     &  3.78\%     \\ \cline{3-5} 
                      &                        & 1120                                                                  & 4.89\%     &  5.94\%     \\ \cline{3-5} 
                      &                        & 1250                                                                  & 6.61\%     &  6.42\%     \\ \cline{3-5} 
                      &                        & 1400                                                                  & 6.18\%     & 7.80\%      \\ \cline{3-5} 
                      &                        & 2800                                                                  & 9.08\%     & 11.65\%    \\ \hline
\multirow{6}{*}{50K}  & \multirow{6}{*}{1.25M} & 300                                                                   & 1.57\%     & 1.43\%     \\ \cline{3-5} 
                      &                        & 600                                                                   & 1.91\%     & 2.11\%     \\ \cline{3-5} 
                      &                        & 1200                                                                  & 2.39\%     & 2.83\%     \\ \cline{3-5} 
                      &                        & 1500                                                                  & 3.74\%     & 3.90\%      \\ \cline{3-5} 
                      &                        & 2500                                                                  & 4.43\%     & 5.23\%     \\ \cline{3-5} 
                      &                        & 3000                                                                  & 4.37\%     & 5.04\%     \\ \hline
\multirow{6}{*}{75K}  & \multirow{6}{*}{2.81M} & 320                                                                   & 1.46\%     & 1.43\%     \\ \cline{3-5} 
                      &                        & 640                                                                   & 2.41\%     & 1.96\%     \\ \cline{3-5} 
                      &                        & 1280                                                                  & 1.89\%     & 2.23\%     \\ \cline{3-5} 
                      &                        & 1600                                                                  & 2.05\%     & 1.87\%     \\ \cline{3-5} 
                      &                        & 3200                                                                  & 3.12\%     & 3.60\%      \\ \cline{3-5} 
                      &                        & 3750                                                                  & 3.44\%     & 4.34\%     \\ \hline
\multirow{6}{*}{100K} & \multirow{6}{*}{5M}    & 320                                                                   & 1.23\%     & 1.30\%      \\ \cline{3-5} 
                      &                        & 640                                                                   & 1.59\%     & 1.16\%     \\ \cline{3-5} 
                      &                        & 1280                                                                  & 2.01\%     & 2.15\%     \\ \cline{3-5} 
                      &                        & 1600                                                                  & 1.93\%     & 1.61\%     \\ \cline{3-5} 
                      &                        & 3200                                                                  & 2.54\%     & 2.76\%     \\ \cline{3-5} 
                      &                        & 5000                                                                  & 3.44\%     & 3.65\%     \\ \hline
\end{tabular}
\caption{\textbf{ER Graphs Results}. The RP (Random Placement) column shows the percentage reduction in toxicity when the peace-bots are placed randomly in the network, while the LI (Lowest Indegree) column shows the percentage reduction when they are placed as incoming neighbours of lowest indegree vertices. The latter almost always outperforms the former.}
\label{tab:er_graph_results}
\end{table}

\begin{table}[h]
\centering
\setlength{\extrarowheight}{1pt} 
\begin{tabular}{|r|r|r|r|r|r|}
\hline
\textbf{\begin{tabular}[c]{@{}c@{}}Real \\ World\\ Graph\end{tabular}} & \textbf{Nodes}         & \textbf{Edges}         & \textbf{\begin{tabular}[c]{@{}c@{}}No. \\ of\\ Peace \\ Bots\end{tabular}} & \multicolumn{1}{c|}{\textbf{RP}} & \multicolumn{1}{c|}{\textbf{LI}} \\ \hline
\multirow{6}{*}{Twitter}                                               & \multirow{6}{*}{100K}  & \multirow{6}{*}{2.28M} & 320                                                                        & 1.26\%     & 0.89\%     \\ \cline{4-6} 
                                                                       &                        &                        & 640                                                                        & 2.08\%     & 0.96\%     \\ \cline{4-6} 
                                                                       &                        &                        & 1280                                                                       & 3.52\%     & 0.68\%     \\ \cline{4-6} 
                                                                       &                        &                        & 1600                                                                       & 3.71\%     & 0.73\%    \\ \cline{4-6} 
                                                                       &                        &                        & 3200                                                                       & 6.10\%      & 1.24\%     \\ \cline{4-6} 
                                                                       &                        &                        & 5019                                                                       & 8.88\%     & 2.49\%     \\ \hline
\multirow{6}{*}{Koo}                                                   & \multirow{6}{*}{275K}  & \multirow{6}{*}{2.33M} & 360                                                                        & 4.66\%     & 6.31\%     \\ \cline{4-6} 
                                                                       &                        &                        & 720                                                                        & 4.16\%     & 5.57\%     \\ \cline{4-6} 
                                                                       &                        &                        & 1440                                                                       & 11.55\%    & 9.34\%     \\ \cline{4-6} 
                                                                       &                        &                        & 1800                                                                       & 11.16\%    & 7.57\%     \\ \cline{4-6} 
                                                                       &                        &                        & 3600                                                                       & 10.15\%    & 9.24\%     \\ \cline{4-6} 
                                                                       &                        &                        & 13751                                                                      & 18.33\%    & 16.44\%    \\ \hline
\multirow{6}{*}{Gab}                                                   & \multirow{6}{*}{72.8K} & \multirow{6}{*}{2.31M} & 320                                                                        & 1.89\%     & 1.65\%     \\ \cline{4-6} 
                                                                       &                        &                        & 640                                                                        & 2.51\%     & 2.78\%     \\ \cline{4-6} 
                                                                       &                        &                        & 1280                                                                       & 3.90\%      & 3.54\%     \\ \cline{4-6} 
                                                                       &                        &                        & 1600                                                                       & 4.61\%     & 4.16\%     \\ \cline{4-6} 
                                                                       &                        &                        & 3200                                                                       & 6.92\%     & 7.32\%     \\ \cline{4-6} 
                                                                       &                        &                        & 3644                                                                       & 7.78\%     & 7.71\%     \\ \hline
\end{tabular}
\caption{\textbf{Real-world Graphs Results:} The RP (Random Placement) column shows the percentage reduction in toxicity when the peace-bots are placed randomly in the network, while the LI (Lowest Indegree) column shows the percentage reduction when they are placed as incoming neighbours of lowest indegree vertices. For Twitter, the RP strategy performs better than the LI strategy, with increasing number of peace-bots; For Koo, the RP strategy performs better for higher number of peace-bots; For Gab, they both are very close.}
\label{tab:real_graph_results}
\end{table}

Table~\ref{tab:er_graph_results} appears to have a clear winner in the LI strategy. This is intuitively easy to explain. All else being ``equal", consider two vertices $u$ and $v$, such that $indeg(u) < indeg(v)$ and 
{\it avgintox(u) = avgintox(v)}. The smaller the number of incoming neighbours, the larger the contribution of the peace-bot to the average incoming toxicity. A peace-bot adds zero toxicity, so the updated values of average incoming toxicity for $u$ and $v$ after the addition of the peace-bot are updated as follows:

\begin{equation}
{\it avgintox(u)} = \frac{ avgintox(u) \cdot indeg(u) + 0 }{ indeg(u) + 1 } - avgintox(u)
\end{equation}

\begin{equation}
{\it avgintox(v)} = \frac{ avgintox(v) \cdot indeg(v) + 0 }{ indeg(v) + 1 } - avgintox(v)
\end{equation}

Since $indeg(u) < indeg(v)$, we expect ${\it avgintox(u)} > {\it avgintox(v)}$ after peace-bot intervention. 

However, the experiments on the real-life datasets tell a different story. For Twitter, the RP strategy performs better than the LI strategy, with increasing number of peace-bots; For Koo, the RP strategy performs better for higher number of peace-bots; For Gab, they both are very close. 

This suggests that the outcome depends considerably on the structure of the network, and we cannot estimate the impact without conducting experiments.
So, when deploying peace-bots on any network in practice, one would have to conduct such experiments, until we get a deeper understanding of the underlying processes which govern user behaviour, as proposed in Section~\ref{sec:discuss}.

%% file: sections/07_discussion.tex
\section{Discussion}
\label{sec:discuss}

We now discuss the limitations of our approach and future work.
\begin{itemize}
\item {\it Assumption of peace-bot connectivity}:~In our experiments, peace-bots were deployed and users were assumed to follow them. In reality, this would require recommending peace-bot accounts to users~\cite{tj2025}, and the effectiveness would depend on how many users accept such recommendations.
\item{\it Dependence on shift distributions}:~The success of peace-bots in reducing average incoming toxicity relies on assumptions about the shift distributions $D_{\textit amp}, D_{\textit atn}, D_{\textit cc}$. If shifts increase relative to the decrease in incoming toxicity, the intervention may fail. Identifying the properties these distributions must satisfy for the intervention to succeed is essential, though not sufficient. Network structure and the placement of user categories also strongly influence the outcome.
\item{\it Why do some users change their behaviour?}~While there was no evidence of homophily, the question as to why some users
change their behaviour and others don't remains. Is this driven by intrinsic user traits, or by their past exposure to toxicity? Insights from social science could help us better understand how individuals respond to toxic content, leading to a more complete picture of the psychological+social individual.
\item{\it Finer models}:~In this paper, we classified users into three categories based on their overall response across all toxicity values. However, users may behave differently across ranges of input toxicity. For example, a user may act as an amplifier for input toxicity in the range [0.23 -0.55], but acts as an attenuator for the rest [0.56 - 1]. This has not been addressed in this work and deserves to be explored. 
\item{\it Network properties}:~Instead of relying on extensive experiments, can we identify properties (if such exist) that might be better predictors of which strategy (random vs. lowest-incoming, etc.) is likely to be more successful for a given network?
\item{\it Holistic models}:~Past research has taken the approach of categorising users as being hateful vs. non-hateful. Whereas this work classifies them as amplifiers, attenuators, or copycats (with possible temporal changes). Combining these two perspectives could yield six user categories, enabling a more comprehensive model and new insights.
\end{itemize}


%% file: sections/08_conclusion.tex
\section{Conclusion}
\label{sec:conclude}

We propose a model to capture the spread of toxicity on online social networks. Based on their behaviour as transformers of toxicity, users may be divided into three categories -- amplifiers, attenuators, and copycats. In the case of Twitter, we found that nearly half of users remain fixed in their respective categories. Our analysis of Twitter, Gab and Koo datasets show that neither spreading activation nor epidemiological models capture the spread of toxicity effectively. Each user applies a shift to the incoming toxicity and the shift is dependent on the user's category and the value of the incoming toxicity.  

Based on these observations, we propose an intervention strategy to mitigate spread of toxicity over time. Our experiments yielded a critical insight: there is no universally optimal strategy for deploying these interventions. While targeting users with the lowest in-degree proved most effective on random networks, this did not hold true for the complex, real-world structures of Twitter, Koo and Gab. On these platforms, a random placement strategy was often more, or equally, effective. \\

These experiments suggest that under certain conditions, it is possible to mitigate toxicity in a social network. If such conditions exist, then the moderators/providers of a social network application can use these techniques to mitigate toxicity. The insights from this work can also inform moderation policy for online platforms.

%% file: sections/10_appendix.tex
\appendix
\section{Appendix}
\subsection{Preliminary Analysis}
\label{sec:appendix_preliminary_analysis}

\begin{figure}[h]
\centering
\includegraphics[width=\columnwidth]{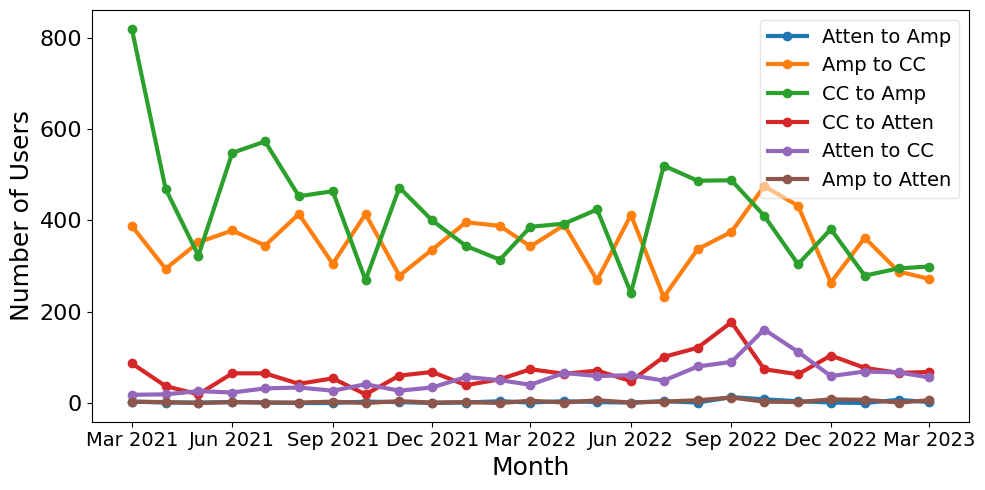}
\caption{User Category change over time in the Koo Dataset}
\label{fig:user_category_change_over_time_koo}
\end{figure}

\begin{figure}[h]
\centering
\includegraphics[width=\columnwidth]{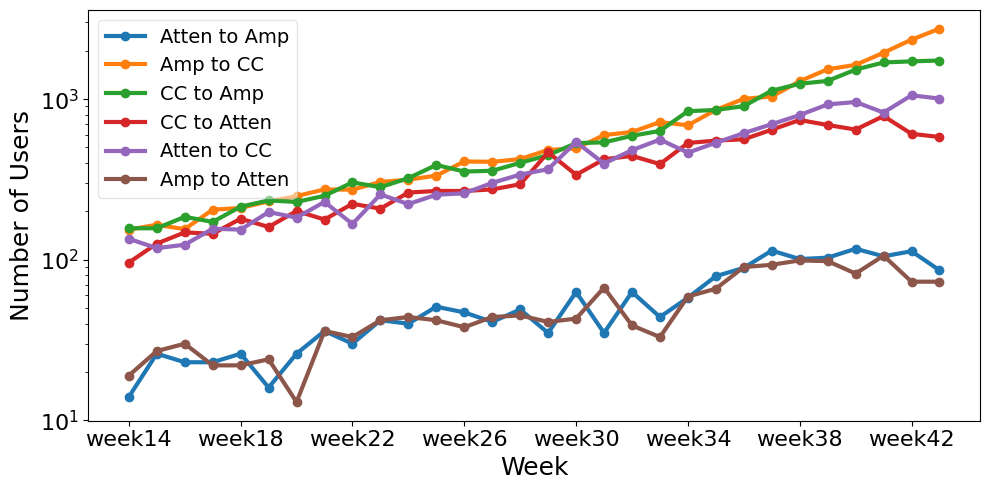}
\caption{User Category change over time in the Twitter Dataset}
\label{fig:user_category_change_over_time_twitter}
\end{figure}


\begin{table}[h]
\begin{tabular}{cccc}
\hline
\textbf{User Category} & \textbf{CopyCat} & \textbf{Attenuator} & \textbf{Amplifier} \\ \hline
\textbf{CopyCat}       & --               & 0.1737              & 0.2826             \\
\textbf{Attenuator}    & 0.1874           & --                  & 0.0347             \\
\textbf{Amplifier}     & 0.2903           & 0.0314              & --                 \\ \hline
\end{tabular}
\caption{User Category Change Probablities}
\label{tab:user_category_change_prob}
\end{table}

